\documentclass[pra,aps,a4paper,onecolumn,11pt,accepted=2021-10-21]{quantumarticle}
\pdfoutput=1
\usepackage{dsfont}
\usepackage{bm}
\usepackage{verbatim}
\usepackage{flushend}
\usepackage{xcolor}
\usepackage{cancel}
\usepackage[utf8]{inputenc}  
\usepackage[T1]{fontenc}     %Output what you want e.g., é, ł, a, ü
\usepackage[british]{babel}  %Do hyphenation according to british english
\usepackage[colorlinks=true, citecolor=blue, urlcolor=blue]{hyperref}  %Hyperlinks (pink, green, blue)
\usepackage{graphicx} % Package to insert exteral figures
\usepackage[normalem]{ulem}
\usepackage[babel]{microtype}  %Improves text justification
\usepackage{amsmath,amssymb,amsthm,bm,amsfonts,mathrsfs,bbm} %Usefull math packages

\usepackage{xspace}  %Useful to add space in macros
\definecolor{wrwrwr}{rgb}{0.3803921568627451,0.3803921568627451,0.3803921568627451}
\definecolor{rvwvcq}{rgb}{0.08235294117647059,0.396078431372549,0.7529411764705882}
\usepackage{pgf,tikz,pgfplots}
\usetikzlibrary{calc}
\usepackage{mathrsfs}
\usetikzlibrary{arrows,snakes}
\usetikzlibrary{positioning}
\usepackage{xcolor}
\usepackage{appendix}
\usepackage{multirow}
\usepackage{array}
\usepackage{bigstrut}
\usepackage{braket}
\usepackage{color}
\usepackage{multirow}
\usepackage{float}
\usepackage[caption = false]{subfig}
\usepackage{xcolor,colortbl}
\usepackage{color}

\newcommand{\be}{\begin{equation}}
	\newcommand{\ee}{\end{equation}}
\newcommand{\ba}{\begin{eqnarray}}
	\newcommand{\ea}{\end{eqnarray}}

\newtheorem{theorem}{Theorem}

\newtheorem{definition}{Definition}
\newtheorem{proposition}{Proposition}

\newtheorem{remark}{Remark}
\newtheorem{lemma}{Lemma}
\definecolor{rvwvcq}{rgb}{0,0,1}
\usepackage{tabu}
\def\>{\rangle}
\def\<{\langle}

\makeatletter
\DeclareRobustCommand{\cev}[1]{%
	{\mathpalette\do@cev{#1}}%
}
\newcommand{\do@cev}[2]{%
	\vbox{\offinterlineskip
		\sbox\z@{$\m@th#1 x$}%
		\ialign{##\cr
			\hidewidth\reflectbox{$\m@th#1\vec{}\mkern4mu$}\hidewidth\cr
			\noalign{\kern-\ht\z@}
			$\m@th#1#2$\cr
		}%
	}%
}
\makeatother

\begin{document}
	
	\title{Quantum Advantage for Shared Randomness Generation}
	
	\author{Tamal Guha}
	\affiliation{Physics and Applied Mathematics Unit, Indian Statistical Institute, 203 B.T. Road, Kolkata 700108, India.}
	
	\author{Mir Alimuddin}
	\affiliation{School of Physics, IISER Thiruvanathapuram, Vithura, Kerala 695551, India.}
	
	\author{Sumit Rout}
	\affiliation{International Centre for Theory of Quantum Technologies (ICTQT), University of Gdańsk, 80-308 Gdańsk, Poland.}
	
	\author{Amit Mukherjee}
	\affiliation{S.N. Bose National Center for Basic Sciences, Block JD, Sector III, Salt Lake, Kolkata 700098, India.}
	
	\author{Some Sankar Bhattacharya}
	\affiliation{Department of Computer Science, The University of Hong Kong, Pokfulam Road, Hong Kong.}
	
	\author{Manik Banik}
	%\email{manik11ju@gmail.com}    
	\affiliation{School of Physics, IISER Thiruvanathapuram, Vithura, Kerala 695551, India.}
	
	\begin{abstract}
		Sharing correlated random variables is a resource for a number of information theoretic tasks such as privacy amplification, simultaneous message passing, secret sharing and many more. In this article, we show that to establish such a resource called shared randomness, quantum systems provide an advantage over their classical counterpart. Precisely, we show that appropriate albeit fixed measurements on a
		shared two-qubit state can generate correlations which cannot be obtained from any possible state on two classical bits. In a resource theoretic set-up, this feature of quantum systems can be interpreted as an advantage in winning a two players co-operative game, which we call the `non-monopolize social subsidy' game. It turns out that the quantum states leading to the desired advantage must possess non-classicality in the form of quantum discord. On the other hand, while distributing such sources of shared randomness between two parties via noisy channels, quantum channels with zero capacity as well as with classical capacity strictly less than unity perform more efficiently than the perfect classical channel. Protocols presented here are noise-robust and hence should be realizable with state-of-the-art quantum devices.

		%Randomness appears both in classical stochastic physics and in quantum mechanics. A number of communication models and other information protocols entail sharing correlated random variables among distant parties. Here we report a computational scenario of shared randomness generation where quantum sources manifest precedence over the corresponding classical counterparts. The advantage is established in a resource theoretic set-up. 
		%In this framework classical mutual information turns out to be a faithful resource quantifier of shared randomness, but fails to characterize all possible resource transitions. 
		%We show that a particular form of shared randomness can be obtained from a finite dimensional bipartite quantum system under the action of free operation, which otherwise is not possible to get from the corresponding classical system. The quantum advantage is manifested in the payoff of a two players co-operative game, namely the `non-monopolize social subsidy' game. Quantum states leading to the desired advantage necessitate non-classicality in the form of quantum discord. On the other hand, while distributing sources of shared randomness between two parties via noisy channels imply that quantum channels with zero capacity and imperfect classical channels with strictly less than unit capacity perform more efficiently than perfect classical channel. These protocols with noisy channels encourage experiments to demonstrate the quantum advantage reported here, with state-of-the-art quantum devices.
	\end{abstract}
	
	%\pacs{03.65.Ta,03.65.Ud, 03.67.Dd}
	%\keywords{}
	
	% 03.65.Ta	Foundations of quantum mechanics;
	% 03.67.Dd	Quantum cryptography and communication security
	% 03.67.Hk	Quantum communication
	% 03.65.Ud	Entanglement and quantum nonlocality
	
	\maketitle
	\section{Introduction}
	Present day quantum technology is getting increasingly sophisticated with the aim to control individual quantum systems, enabling them in different practical tasks that otherwise are not possible in classical world. This approach already finds several practical applications, such as secure communication \cite{Jiang2007,Dixon2008,Wengerowsky2019,Yin2020}, quantum imaging \cite{Tan2008,Schneider2017,Asban2019,Gregory2020}, quantum metrology \cite{Roos2006,Appel2009,Giovannetti2011,Kurizki2015}, and more excitingly promises opportunities for Near-Term Quantum Computing Systems \cite{Li2001,Larsen2019,Kandala2019,Flhmann2019,Arute2019, GarcaPatrn2019}. Thus, it is important to explore and identify more and more instances where quantum theory can exhibit advantage over the corresponding classical systems. %Such advantages, however, are notoriously hard to find and even harder to prove. For instance the set of functions computable quantum mechanically is the same as the set computable in classical physics. \textcolor{blue}{However, using non-classical features of quantum world it is possible to have quantum algorithms that solve some problems efficiently than the classical ones \cite{Shor1994, Grover1996, Simon1997}. } 
	In this work, we report such a novel quantum advantage. We consider the computational scenario of generating correlated random variables between distant parties, also known as {\it shared randomness}. 
	
	Shared randomness (SR) is known to be an important resource in a number of applications, {\it viz} privacy amplification \cite{Bennett1988,Bennett1995,Newman1996}, simultaneous message passing \cite{Babai1997}, secret sharing and secret key generation protocol \cite{Gavinsky2012,Ahlswede1993}, classical simulation of quantum nonlocal statistics \cite{Brassard1999,Toner2003,Bowles2015}, Bayesian game theory \cite{Aumann1987,Brunner2013,Roy2016,Banik2019}, and communication complexity \cite{Canonne2017}. Among spatially separated parties, shared randomness can not be established free of cost. It requires the distant parties to have access to noiseless communication channels, which, in Shannon theory, are considered to be expensive \cite{Shannon1948}. Alternatively, one can ask whether sharing multipartite quantum systems provide any advantage over the correlated classical systems for shared randomness generation or not. A similar question can also be asked concerning the advantage of using noisy quantum communication channels over classical ones. In this work, we find affirmative answers to both of these questions by identifying new instances where quantum theory yields provable advantage over its classical counterpart. Importantly, the quantum advantage sustains even in presence of noise and hence is achievable with the present day imperfect quantum devices.
	
	To demonstrate the advantage of using quantum sources, we take resort to the language of resource theory. In the recent past, researchers in quantum information community have successfully applied this framework to identify, characterize, and quantify different useful resources \cite{Braunstein2005,Bartlett2007,Horodecki2009,Modi2012,Brandao2013,Grudka2014,Rivas2014,Veitch2014,Gallego2015,Winter2016,Chitambar2019,Wolfe2020}. Such a framework is operationally motivated. Firstly, it identifies the free states that are {\it useless} for performing certain tasks and specifies the free operations that are unable to produce any resource from free states and hence are allowed to be implemented at no cost. Given these sets of free states and operations, the framework aims to find the resource conversion conditions (either necessary or sufficient, sometimes both), commonly phrased as monotones, that characterize possible transformations among the resource states under free operations.

	In this article, we consider the resource theory of shared randomness. 
	%case of SR, classical mutual information turns out to be a \textit{faithful} quantifier of the resource which takes zero value for all the free states while nonzero for any resourceful state. 
	At the outset, it is worth mentioning that our framework is distinct from the well known resource theory of local operations with shared randomness (LOSR) \cite{Schmid2020, Schmid2020a}. In the present work, shared randomness is not considered as a free resource, which is the case in the resource theory of LOSR. Here, we aim to quantify the resource for generating shared randomness between distant parties by performing local operations on their subsystems. In that respect, the works of Toner {\it et al.}\cite{Toner2003} and Bowles {\it et al.}\cite{Bowles2015} are noteworthy, where it has been shown that nonlocal correlations obtained from a bipartite entangled state can be simulated with shared randomness when assisted  with classical communication. While in \cite{Toner2003} it requires infinite amount of shared randomness, the authors in \cite{Bowles2015} propose a simulation of nonlocal states with finite shared randomness and finite communication. The present article establishes the utility of shared randomness even outside the nonlocality paradigm. First, we identify the set of correlations that can be obtained from a shared 2-faced classical coin (henceforth called {\it two-2-coin}) under the free local operations. Secondly we observe that, within the proposed resource theoretic framework, every two-2-coin state can be freely obtained from its quantum analogue, namely the two-2-quoin which corresponds to a two-qubit quantum system with Hilbert space $\mathbb{C}^2\otimes\mathbb{C}^2$. Lastly, the quantum advantage is established by a set inclusion relation, which involves identifying two-d-coin states that can be obtained from a two-2-quoin under free operation but cannot be obtained from any two-2-coin state. 
	
	We also show that the quantum advantage for generating shared randomness translates to higher success probability of winning a two-player co-operative game, namely the `non-monopolize social subsidy' game with quantum resource, when compared to that of corresponding classical strategies. More precisely, the players can achieve optimal payoff when assisted with two-2-quoin states, whereas their payoffs remain suboptimal with two-2-coin states. Further, we show that {\it better than classical} payoff necessitates use of two-2-quoin states with non-zero discord -- an intriguing non-classical feature present in bipartite quantum systems even when the states are not entangled \cite{Ollivier2001,Henderson2001}. We then consider the scenario where one wishes to establish shared randomness with a distant party. We show that a quantum channel can exhibit advantage over the corresponding classical channels. Such an advantage is quite remarkable, as there exist {\it no-go} results \cite{Holevo1973,Frenkel2015} that limit the utilities of quantum systems as information carrier. Recall that in Shannon theory, efficacy of a classical channel is characterized by its capacity, quantified as the mutual information optimized over probability distributions of the input variables \cite{Shannon1948}. In quantum scenario, different quantities of interest are used to characterize the utility of a quantum channel. For instance, while quantum capacity of a quantum channel denotes the highest rate of transmitting quantum information \cite{Lloyd1997,Shor2002,Devetak2005}, its classical capacity \cite{Holevo1998,Schumacher1997} characterizes utility of transferring classical information. In shared randomness distribution, a quantum channel can show advantage over a classical channel even when its classical capacity is much less than that of the classical channel. At this point, it seems natural to think that such advantage requires the noisy quantum channels to possess non-zero quantum capacity. However, it turns out that the quantum advantage persists even when the quantum channel has zero quantum capacity. Evidently, these instances of noise robust advantage of quantum strategies should be realizable with the state-of-the-art quantum devices.        
	
	\section{Results}
	\subsection{Resource theory of shared randomness} 
	The framework of resource theory provides a novel approach to quantify the resources of shared randomness. The generic framework of any resource theory characterizes the followings: the class of free states or non-resources, the set of free operations, and resource conversion conditions (either necessary or sufficient, sometimes both) that are commonly phrased as monotones \cite{Brandao2015}.
	\subsubsection*{Free resources} A source of shared randomness is specified by a bipartite probability distribution $P(\mathcal{X},\mathcal{Y})\equiv\{ p(x,y)~|~x\in\mathcal{X},~y\in\mathcal{Y}\}$, where $\mathcal{X}$ and $\mathcal{Y}$ are the parts of the shared variable accessible by spatially separated parties Alice and Bob, respectively. Probability distributions of the product form $P(\mathcal{X},\mathcal{Y})=P(\mathcal{X})Q(\mathcal{Y})$ are considered as free resources/states as each of the shared variables follows an independent probability distribution and consequently information of one does not provide any knowledge about the other. Unlike the resource theories of quantum entanglement \cite{Horodecki2009} or quantum coherence \cite{Winter2016} the set $\mathcal{F}_{SR}$ of free states does not form a {convex set in this case}. 
	\begin{figure}[t!]
		\centering
		\includegraphics[width=7cm,height=5.0cm]{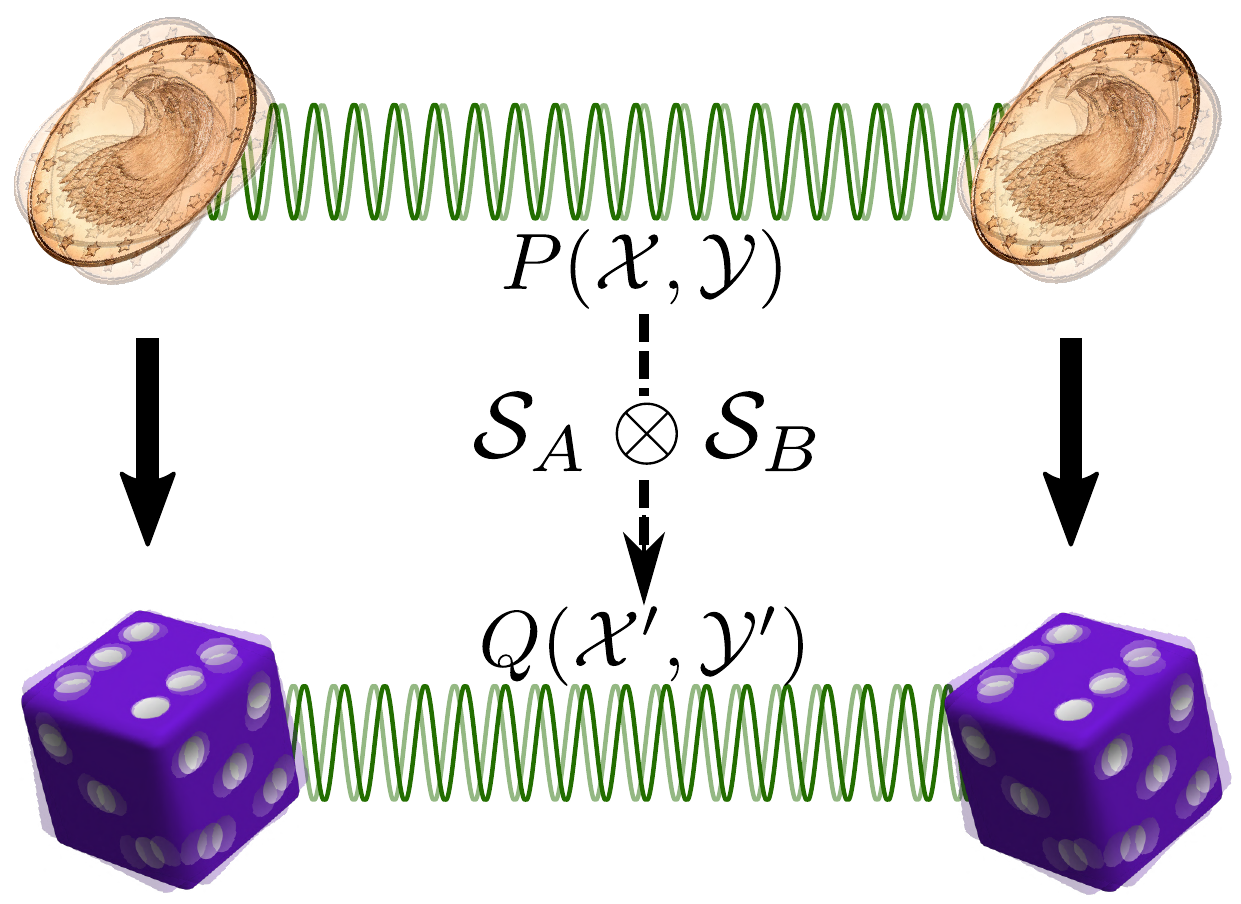}
		\caption{[Color on-line] \textbf{Resource theory of shared randomness processing.} By performing free operations (local stochastic operations) on two-2-coin states $\mathfrak{C}(2)$ one can obtain only a proper subset $\mathfrak{S}_C(2\mapsto d)$ of two-d-coin state space $\mathfrak{C}(d)$. Such a transformation can never increase classical mutual information of the coin state. For instance, the transformation $\mathcal{C}_{1/2}(2)\mapsto\mathcal{C}_{1/6}(6)$ is not allowed under free operations, where $\mathcal{C}_{1/6}(6):=1/6\sum_{f=1}^6\mathbf{ff}\in\mathfrak{C}(6)$.}\label{fig1}
	\end{figure}
	
	In an operational theory, shared randomness between Alice and Bob can be obtained from a shared bipartite system by performing local measurement on their respective parts. The state space of such a system, in a convex operational theory, is given by $\Omega_A\otimes\Omega_B$, where $\Omega_K$ be the convex compact marginal state space embedded in some real vector space $V_K$; $K\in\{A,B\}$ \cite{Hardy2001,Barrett2007,Chiribella2011}. For instance, the state space of $d$-level classical system is the $d$-simplex embedded in $\mathbb{R}^{d-1}$, whereas for $d$-level quantum system it is $\mathcal{D}(\mathbb{C}^d)\subset\mathbb{R}^{d^2-1}$; $\mathcal{D}(\mathcal{H})$ denotes the set of density operators acting on the Hilbert space $\mathcal{H}$ associated with the system. While considering the state space for a composite system by taking tensor product of component state space of the subsystems, it is important to note that the choice of tensor product is unique for simplex, which is not the case for other convex sets \cite{Namioka1969,Barker1976,Barker1981,Aubrun2019}.         
	\subsubsection*{Free operations} The set of free operations for SR consists of all possible local product operations $L_A\otimes L_B$ applied by Alice and Bob on their respective parts of the joint system. For classical systems, such operations are most generally described by tensor product of local stochastic matrices $\mathcal{S}_A\otimes\mathcal{S}_B$, where $\mathcal{S}_A$ maps Alice's local probability vector $P(\mathcal{X})$ into a new probability vector $P^\prime(\mathcal{X}^\prime)$ and $\mathcal{S}_B$ does the similar on Bob's part. Note that cardinality of $\mathcal{X}$ and $\mathcal{X}^\prime$ can be different in general (see Fig.\ref{fig1}). In the quantum scenario, the allowed operations are local unitary operations and/or local measurements generally described by a positive operator valued measure (POVM) \cite{Kraus1983}. At this point, a comparison with the resource theory of quantum entanglement is worth mentioning. In entanglement theory classical communication is considered as free, but it bears a cost in the present scenario as it can create a non-product joint distribution, {\it i.e.}, a resourceful state, starting from a product one. In any operational theory, if Alice and Bob initially share a joint state $\omega_{AB}\in\Omega_A\otimes\Omega_B$ of the product from, \textit{i.e.}, $\omega_{AB}=\omega_{A}\otimes\omega_{B}$, then a free operation on it can never result in an SR resource between them.
	\subsubsection*{Resource monotone} A necessary condition of state conversion from a distribution $P(\mathcal{X,Y})$ to another $Q(\mathcal{X}^\prime,\mathcal{Y}^\prime)$ is given by $I(Q)\le I(P)$, where $I(P)$ is the classical mutual information defined as $I(P):=H(\mathcal{X})+H(\mathcal{Y})-H(\mathcal{X,Y})$, with $H(\mathcal{X})$ being the Shannon entropy, $H(\mathcal{X}):=-\sum_{x\in\mathcal{X}}p(x)\log_2p(x)$. Importantly, mutual information is a faithful resource quantifier, as it takes zero value for every free state while non-zero for all the resourceful states. In the subsequent section, however, we will see that it can not sufficiently characterize the possible resource conversions.
	\begin{figure}[t!]
		\centering
		\includegraphics[width=7.5cm,height=4.5cm]{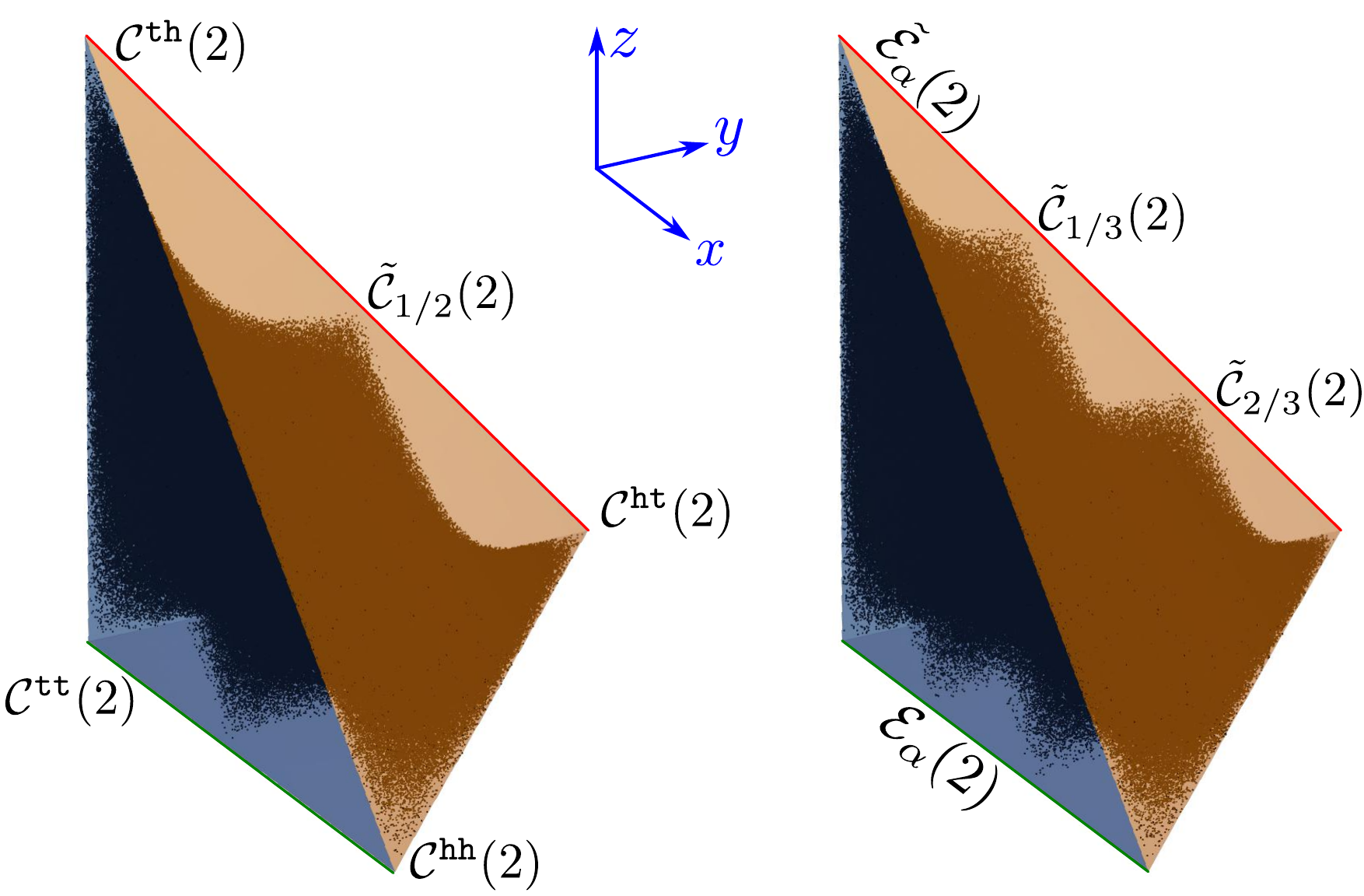}
		\caption{[Color on-line] \textbf{Two-2-coin state space $\mathfrak{C}(2)$.} All the four vertices are the free states. Green (red) line denotes the $\alpha$-correlated ($\alpha$-anti-correlated) edges. The remaining four edges consist of free states only. Dots in the left [right] figure denote the states obtained from $\mathcal{C}_{1/2}(2)$ [$\mathcal{C}_{1/3}(2)$] by applying randomly generated local stochastic maps. Action of such maps on $\alpha$-correlated (or $\alpha$-anti-correlated) edge generate the whole state space $\mathfrak{C}(2)$ (see Lemma \ref{lemma1}).}\label{fig2}
	\end{figure}
	\subsubsection*{Two-2-coin state space} Consider that Alice and Bob share a pair of 2-faced classical coins (two-2-coin), \textit{i.e.}, $\mathcal{X}\equiv\{\mbox{head}(\mathtt{h}),$ $\mbox{tail}(\mathtt{t})\}\equiv\mathcal{Y}$. A generic state of this system is described by a column vector $\mathcal{C}(2)\equiv(p(\mathtt{hh}),p(\mathtt{ht}),p(\mathtt{th}),p(\mathtt{tt}))^\intercal\in\mathfrak{C}(2)$; with $\mathfrak{C}(2)$ denoting the set of all two-2-coin states. A state $\mathcal{C}(2)\equiv(x,y,z,1-x-y-z)^\intercal$ is isomorphic to the vector $\mathcal{V}\equiv(x,y,z)^\intercal\in\mathbb{R}^3$ with $x,y,z\ge 0~\&~x+y+z\le1$, forming a convex subset $\mathbf{T}$ in the positive octant (see Fig.\ref{fig2}). All the four vertices ($0$-faces) $\mathcal{C}^{\mathtt{hh}}(2)$, $\mathcal{C}^{\mathtt{ht}}(2)$, $\mathcal{C}^{\mathtt{th}}(2)$, and $\mathcal{C}^{\mathtt{tt}}(2)$ are free states. We call the states $\mathcal{C}_{\alpha}(2):=(\alpha,0,0,1-\alpha)^\intercal\equiv\alpha~\mathtt{hh}+(1-\alpha)\mathtt{tt}$ as $\alpha$-correlated. Whenever $\alpha\notin\{0,1\}$, $\mathcal{C}_{\alpha}(2)$ contains shared randomness even though they are obtained by convex mixing of two free states, hence implies non-convexity of $\mathcal{F}_{SR}$. The $\alpha$-correlated states live in one of the edges ($1$-faces) of $\mathbf{T}$ and we call it $\alpha$-correlated edge, which will be denoted as $\mathcal{E}_{[\alpha]}(2):=\{\mathcal{C}_{\alpha}(2);~\alpha\in[0,1]\}$. Under free operations, this edge can be transferred into the $\alpha$-anti-correlated edge $\tilde{\mathcal{E}}_{[\alpha]}(2)\equiv\{\tilde{\mathcal{C}}_{\alpha}(2):=\alpha~\mathtt{ht}+(1-\alpha)\mathtt{th};~\alpha\in[0,1]\}$. In fact, every $\mathcal{C}_{\alpha}(2)$ is connected to the corresponding $\tilde{\mathcal{C}}_{\alpha}(2)$ by local permutation, a free operation that keeps the mutual information invariant. The remaining four edges of $\mathbf{T}$ contain only free states. Except these states, no other state residing on any of the four $2$-faces of $\mathbf{T}$ is free. However, the volume ($3$-face) of $\mathbf{T}$ contains both free and resource states.
	
	Consider a state $\mathcal{C}_{\Delta}(2):=\left(1/3,0,1/3,1/3 \right)^\intercal$ residing on one of the $2$-faces of $\mathbf{T}$. The state $\mathcal{C}_{\Delta}(2)$ can be obtained from  $\mathcal{C}_{1/2}(2)$ under free operation. The two possible free operations allowing this transformation are given by,
	$$\left\lbrace \begin{pmatrix} 0 & 2/3 \\1 & 1/3 \end{pmatrix}\otimes\begin{pmatrix} 1/3 & 1 \\2/3 & 0 \end{pmatrix};~~ 
	\begin{pmatrix} 2/3 & 0 \\1/3 & 1 \end{pmatrix}\otimes\begin{pmatrix} 1 & 1/3 \\0 & 2/3 \end{pmatrix}\right\rbrace. $$
	The reverse transformation $\mathcal{C}_{\Delta}(2)\mapsto\mathcal{C}_{1/2}(2)$ is not possible under free operations as the former has lesser mutual information than the latter. Importantly, such a transformation may not be possible even if the initial state has more mutual information than the targeted one. For instance, none of the states $\mathcal{C}_{\alpha}(2)$ can be obtained from $\mathcal{C}_{1/2}(2)$ whenever $\alpha\not\in\left\lbrace0,1/2,1\right\rbrace $, though $I(\mathcal{C}_{1/2}(2))\ge I(\mathcal{C}_{\alpha}(2))$, with strict inequality holding for $\alpha\in[0,1/2)\cup(1/2,1]$. It establishes insufficiency of mutual information in characterizing the possible state conversions. It furthermore proves non-convexity of the set of states obtained from a given resource under the free operations.  
	
	\subsection{Quantum advantage} In this section we will present our main result which establishes quantum advantage in shared randomness processing. To this aim, we first introduce a quantum analogue of the two-2-coin.  
	\subsubsection{Two-2-quoin state space} The quantum analogue of two-2-coin state, which we call two-2-quoin and denoted as $\mathcal{Q}(2)$, corresponds to the states of a two-qubit quantum system. The state space is given by $\mathfrak{Q}(2)\equiv\mathcal{D}(\mathbb{C}^2_A\otimes\mathbb{C}^2_B)$, where subsystems $A$ and $B$ are held by Alice and Bob, respectively. From the two-2-quoin states, Alice and Bob can prepare any state of $\mathfrak{C}(2)$ by applying local POVMs on their respective parts of the joint system. Therefore, the former can always replace the latter for any shared randomness processing task. From these shared classical and quantum 2-level coins, one can obtain shared d-level classical coin states by performing suitable stochastic operations and measurements, respectively. The following proposition establishes quantum advantage in generating shared randomness with higher outcomes.
	\begin{proposition}\label{prop1}
		Let $\mathfrak{S}_C(2\mapsto d)$ denote the set of two-d-coin states in $\mathfrak{C}(d)$ that are freely simulable (\textit{i.e.}, can be obtained under allowed free operations) with states from $\mathfrak{C}(2)$. Similarly, $\mathfrak{S}_Q(2\mapsto d)$ denotes the subset of $\mathfrak{C}(d)$ freely simulable from states in $\mathfrak{Q}(2)$. It holds that 
		$\mathfrak{S}_C(2\mapsto d)\subset\mathfrak{S}_Q(2\mapsto d)$, for $d>2$.
	\end{proposition}
	\noindent
	{\it Sketch of the proof}.
	First we observe that both the sets $\mathfrak{S}_C(2\mapsto d)$ and $\mathfrak{S}_Q(2\mapsto d)$ are non-convex, and for any state $\mathcal{C}(2\mapsto d)\in\mathfrak{S}_J(2\mapsto d)$, with $J\in\{C,Q\}$, we have $I(\mathcal{C}(2\mapsto d))\le1$. Then we argue that $\forall~\mathcal{C}(2\mapsto d)\in\mathfrak{S}_C(2\mapsto d)$, it also lies in $\mathfrak{S}_Q(2\mapsto d)$. Note that a $2-$coin state $\mathcal{C}(2)\equiv(p,q,r,1-p-q-r)^\intercal$ can be obtained from a two-qubit state, $\rho_{AB}=p~|00\rangle\langle00|+ q~|01\rangle\langle01|+ r~|10\rangle\langle10|+ (1-p-q-r)~|11\rangle\langle11|$ by performing local measurement in computational basis. Furthermore, corresponding to every $2\times d$ stochastic matrix, applied locally on $\mathcal{C}(2)$ there is a $d-$outcome POVM acting locally on the part of $\rho_{AB}$, which implies that $\mathfrak{S}_C(2\mapsto d)\subseteq\mathfrak{S}_Q(2\mapsto d)$. Proof of the strict set inclusion relation is deferred till the end of Theorem \ref{theo1} and Theorem \ref{theo2} (see Remark \ref{remark1}). Rather, we now show that the strict set inclusion can be rendered as quantum advantage in a practical two-player game. 
	%	\begin{widetext}
	
	%\Blindtext
	%\end{widetext}
	\subsubsection{Non-monopolizing social subsidy game} The game $\mathbb{G}(n)$ involves two employees Alice \& Bob working in an organization and $n$ different restaurants $r_1,\cdots,r_n$. On every working day, each of the employees buys beverage from the restaurant chosen at her/his will. The organization has a reimbursement policy to pay back the beverage bill. For this purpose, each day's bill is accounted for a long time to calculate the probability $P(ij)$ of Alice visiting $r_i$ restaurant and Bob $r_j$ restaurant. Events $(ij)$ where each employee ends up in different restaurants $(i\ne j)$ are considered for reimbursement / payoff\footnote{This condition mimics the physical distancing norm which people need to follow during the unfortunate pandemic of COVID-19.}. Now it may be the case that the employees pick their favorite restaurants which happen to be different and become regular visitors. But this will leave the other restaurants out of business. To circumvent this situation a sub-clause is added to the subsidy rule which says that the payoff will be defined as $\$\mathcal{R}(n)= \$\min_{i\ne j}P(ij)$, maximizing over all possible strategies with a source of shared randomness with fixed local level, allowed by any physical theory. We further assume that the per day expense for each of the employees is $\$1$. Since the reimbursement policy encourages total trade to be distributed among all the restaurants, we call it `non-monopolizing subsidy' rule. The employees are non-communicating and possess a bipartite state with subsystems described by two-level systems, independent of the number of restaurants. They can choose local strategies from the set of free operations. Following result bounds their achievable payoff.
	\begin{figure*}[ht]
		\centering
		\includegraphics[width=1\textwidth]{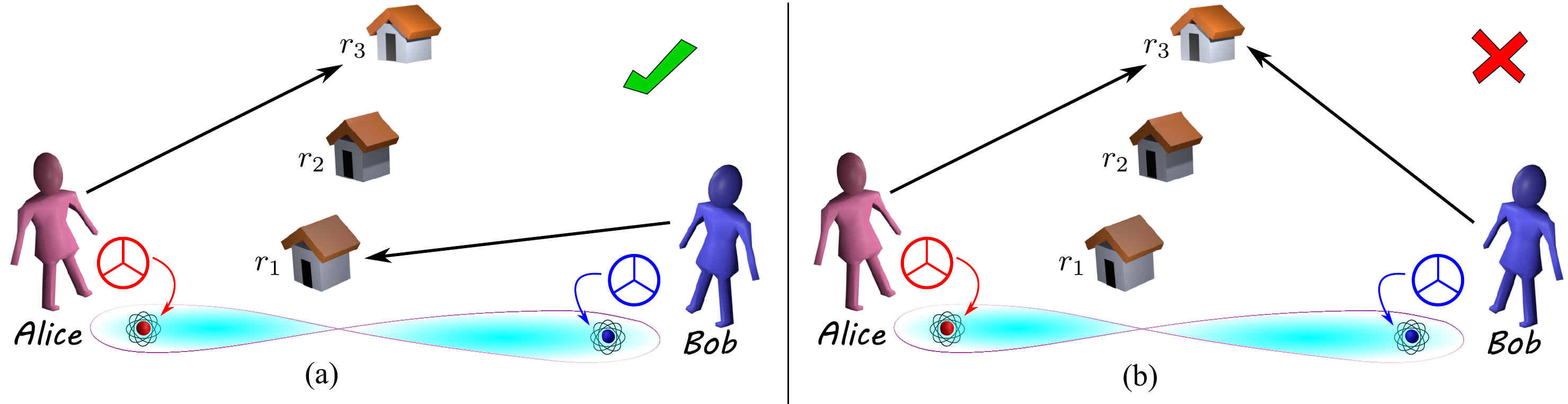}
		\caption{[Color on-line] \textbf{Non-monopolizing social subsidy game.} The choice of restaurants, as in figure (a), is permissible by the organization to obtain a subsidy on the beverage costs of the employees, while the situation depicted in figure (b) will not be entertained. Although a two-2-coin state is unable to accomplish all possible combinations similar to (a), a shared two-2-quoin can do so.}\label{fig3}
	\end{figure*}
	\begin{proposition} The maximum payoff achieved in the game $\mathbb{G}(n)$ by two spatially separated employees is bounded from above and below by the following expression-\\   
		$$\frac{1}{n^2}\le\mathcal{R}(n)\le\frac{1}{n(n-1)}$$.
	\end{proposition}
	\begin{proof}
		The lower bound is achieved, on an average, when a uniformly randomized local strategy is followed by each of the employees. For maximal payoff, there are $n(n-1)$ different cases where the employees' bills get reimbursed. Since minimum probability of these events will be considered for reimbursement, the optimal payoff will be achieved if they choose these cases with equal probability, \textit{i.e.}, with probability $\frac{1}{n(n-1)}$. Note that the payoff of the employees will be zero if both the employees decide to go to the same restaurant every day. Thus, the lower bound in the above proposition assumes rational employees who want to maximize their payoff.
	\end{proof}
	At this point, we define a specific kind of bipartite shared randomness, namely \textit{not-$\alpha$-correlated} coin, which in a special case saturates the upper bound of the payoff in the $\mathbb{G}(n)$ game.
	\begin{definition}\label{def}
		A two-$d$-coin state is said to be `not-$\alpha$-correlated' if $p(\mathbf{ff})=0\text{ and }p(\mathbf{ff^\prime})\ne0,~\forall~\mathbf{f},\mathbf{f^\prime}\in\{1,\cdots,d\},\text{ and }\mathbf{f}\ne\mathbf{f^\prime}$.
	\end{definition}
	In the rest of the manuscript, we will depict them as $\mathcal{C}_{\ne\alpha}(d)\in\mathfrak{C}(d)$.
	The maximum achievable payoff in $\mathbb{G}(n)$ is assured if the employees share a particular not-$\alpha$-correlated coin state $\mathcal{C}^{eq}_{\ne\alpha}(n)$, where $p(\mathbf{ff^\prime})=1/n(n-1),~\forall~\mathbf{f},\mathbf{f^\prime}\in\{1,\cdots,n\},~\&~\mathbf{f}\ne\mathbf{f^\prime}$. What follow next are the two Lemmas regarding simulability of different sets of classical coin states using free operations. 
	\begin{lemma}\label{lemma1}
		Under the action of free operations, any coin state of $\mathfrak{C}(2)$ can be obtained from the $\alpha$-correlated edge $\mathcal{E}_{[\alpha]}(2)$, {\it i.e.}, $\mathcal{E}_{[\alpha]}(2)$ freely simulates the state space $\mathfrak{C}(2)$.
	\end{lemma}
	\begin{proof}
		A state $\mathcal{C}(2)\in\mathfrak{C}(2)$ can most generally be expressed as,
		\begin{eqnarray}\label{a1}
			\mathcal{C}(2)&=&\left(x,y,z,1-x-y-z\right)^\intercal,\\
			0 \leq x \leq 1;&&	0\leq y\leq 1-x;~~~~~	0\leq z\leq 1-x-y.\nonumber
		\end{eqnarray}
		The range of $y$ is determined by the value of $x$, {\it i.e.}, $\forall~x\in[0,1]$ the value of $y$ lies within $[0,1-x]$. Similarly, the range of $z$ is specified by $x$ and $y$. Even though the variables specify each other's range, their values are mutually random, {\it i.e.}, the variables fix the range of each other but not the exact value. We wish to show that by applying local stochastic operations on $\mathcal{C}_\alpha(2)\in\mathcal{E}_{[\alpha]}(2)$ Alice and Bob can prepare any vector of the form of Eq.(\ref{a1}). We therefore can write 
		\begin{eqnarray}\nonumber
			&&\mathcal{K}:=\mathcal{S}^{2\mapsto2}_A\otimes\mathcal{S}^{2\mapsto2}_B~\times~\mathcal{C}_\alpha(2)\\
			&&=\begin{pmatrix} a_{1} & 1-a_{1} \\a_{2} & 1-a_{2}\end{pmatrix}^\intercal\otimes\begin{pmatrix} b_{1} & 1-b_{1} \\b_{2} & 1-b_{2}\end{pmatrix}^\intercal\times\begin{pmatrix} \alpha  \\0 \\0 \\1-\alpha\end{pmatrix}\nonumber\\
			&&=\begin{pmatrix} a_{1}b_{1}\alpha+a_{2}b_{2}(1-\alpha)~~[:=k_1]  \\a_{1}\alpha + a_{2}(1-\alpha)-k_1~[:=k_2] \\b_{1}\alpha+b_{2}(1-\alpha)-k_1~~[:=k_3] \\1-\sum_{i=1}^3k_i~~~~~~~~~~~~~~~\end{pmatrix}=\begin{pmatrix} k_1  \\k_2 \\k_3 \\1-\sum k_i\end{pmatrix},\nonumber
		\end{eqnarray}
		where $a_{1},a_{2},b_{1},b_{2}\in[0,1]$. Since action of a local stochastic matrix $\mathcal{S}_A\otimes\mathcal{S}_B$ on $\mathcal{C}_{\alpha}(2)$ always result in a probability vector, therefore constraints as of Eq.(\ref{a1}) among $k_1,k_2,$ and $k_3$ are always satisfied. Now, for every fixed values of $a_{2},b_{2}\in[0,1]$, $\exists~\alpha,a_{1},b_{1}\in[0,1]$ s.t. $k_1$ can take all values in $[0,1]$. Since the values of $a_{2}$ and $b_{2}$ can be chosen randomly, they are independent of each other and also $k_1$ is independent of them. Consequently, $k_2$ and $k_3$ are independent of $k_1$ and also of each other. This completes the proof.
	\end{proof}
	\begin{lemma}\label{lemma2}
		None of the coin states $\mathcal{C}_{\neq\alpha}(n)$ are freely simulable from $\mathcal{E}_{[\alpha]}(2)$, whenever $n> 2$.
	\end{lemma}
	\begin{proof}
		A generic stochastic operation $\mathcal{S}^{2\mapsto n}$ mapping a two-level probability vector into an $n$ level probability vector is of the form 
		$$\begin{pmatrix} u_{11} &u_{21}&\cdots& 1-\sum_{i=1}^{n-1}u_{i1} \\u_{12} &u_{22}&\cdots& 1-\sum_{i=1}^{n-1}u_{i2}\end{pmatrix}^\intercal,$$
		where $u_{ij}\in[0,1]$ and $\sum_{i=1}^{n-1}u_{ij}\le1$. Action of local operations by Alice and Bob on their respective parts of the coin state $\mathcal{C}_\alpha(2)$ yield a two-$n$-coin state,
		\begin{eqnarray}\nonumber
			\mathcal{C}(n)&=&\mathcal{S}^{2\mapsto n}_A\otimes\mathcal{S}^{2\mapsto n}_B~\times~\mathcal{C}_\alpha(2)\\\nonumber
			&=&\begin{pmatrix} a_{11} &a_{21}&\cdots& 1-\sum_{i=1}^{n-1}a_{i1} \\a_{12} &a_{22}&\cdots& 1-\sum_{i=1}^{n-1}a_{i2}\end{pmatrix}^\intercal\otimes\\\nonumber
			&&\begin{pmatrix} b_{11} &b_{21}&\cdots& 1-\sum_{i=1}^{n-1}b_{i1} \\b_{12} &b_{22}&\cdots& 1-\sum_{i=1}^{n-1}b_{i2}\end{pmatrix}^\intercal\times\begin{pmatrix} \alpha  \\0 \\0 \\1-\alpha\end{pmatrix}.
		\end{eqnarray}
		Whenever $\alpha\in\{0,1\}$, the initial state is free and hence the final one. To get the final state as $\mathcal{C}_{\neq \alpha}(n)$ (see Definition\ref{def}), we require $\alpha a_{i1}b_{i1}+(1-\alpha)a_{i2}b_{i2}=0,~\forall~i\in\{1,...,n\}$. Since $\alpha\in(0,1)$, therefore $a_{ij}b_{ij}=0,~\forall~i\in\{1,...,n\}~\&~\forall~j\in\{1,2\}$. Presence of anti-correlated terms in $\mathcal{C}_{\neq \alpha}(n)$ demands, $\alpha a_{i1}b_{k1}+(1-\alpha)a_{i2}b_{k2}\neq0,~\forall~ i,k\in\{1,...,n\}~\&~i\neq k$. Therefore, for every $(i,k\neq i)$ pair $\exists$ at-least one $j\in\{1,2\}$ s.t. $a_{ij}b_{kj}\neq0\implies a_{ij}\neq0$ and $b_{kj}\neq0$. Similarly, for the corresponding reverse pair, $(k,i\neq k)~\exists$ at-least one $j'\in\{1,2\}$ s.t. $a_{kj'}b_{ij'}\neq0\implies a_{kj'}\neq0$ and $b_{ij'}\neq0$. Now $j$ and $j'$ should be different, otherwise a correlated term of the resulting coin state will become non-vanishing. Since $j,j'\in\{1,2\}$, the requirement $j\neq j'$ can not be satisfied whenever $i,k\in\{1,...,n\}$, with $n>2$. This completes the proof.
	\end{proof}
	These two Lemmas lead us to the following result, describing the limitation of the shared classical coins in achieving the maximum payoff in the game $\mathbb{G}(n)$.
	\begin{theorem}\label{theo1}
		Given any coin state from $\mathfrak{C}(2)$, the payoff $\mathcal{R}(n)$ is always suboptimal for all $n>2$.
	\end{theorem}
	\begin{proof}
		Contrary to the hypothesis, let us assume that there exist a two-2-coin state $\mathcal{C}^n_{win}(2)$ that provides perfect success in $\mathbb{G}(n)$. Since perfect success of $\mathbb{G}(n)$ requires the two-$n$-coin state $\mathcal{C}^{eq}_{\neq\alpha}(n)$, this implies that $\mathcal{C}^{eq}_{\neq\alpha}(n)$ can be obtained from $\mathcal{C}^n_{win}(2)$ under free operation. Invoking Lemma \ref{lemma1} we can say that the state $\mathcal{C}^{eq}_{\neq\alpha}$ can be obtained freely from $\mathcal{E}_{[\alpha]}(2)$. This, however, contradicts Lemma \ref{lemma2}.
	\end{proof}
	At this point one can ask for maximum payoff $\mathcal{R}^{\mathfrak{C}(m)}_{\max}(n)$ that can be achieved in $\mathbb{G}(n)$ given an assistance from $\mathfrak{C}(m)$. This turns out to be an optimization problem. Given a two-$m$-coin, $\mathcal{C}(m)\equiv\left(p(\mathbf{11}),\cdots,p(\mathbf{1m}),\cdots,p(\mathbf{mm}) \right)^\intercal$ Alice and Bob can obtain some two-$n$-coin states $\mathcal{C}(n)\equiv\left(q(\mathbf{11}),\cdots,q(\mathbf{1n}),\cdots,q(\mathbf{nn}) \right)^\intercal$ by applying local stochastic maps (free operation), \textit{i.e.}, $\mathcal{C}(n)=\mathcal{S}^{m\mapsto n}_A\otimes\mathcal{S}^{m\mapsto n}_B\cdot \mathcal{C}(m)$. We therefore have
	\begin{eqnarray}\nonumber
		\mathcal{R}^{\mathfrak{C}(m)}_{\max}(n)&=&\underset{\underset{\mathcal{S}^{m\mapsto n}_A\otimes\mathcal{S}^{m\mapsto n}_B}{\mathcal{C}(m)\in\mathfrak{C}(m)}}{\mbox{maximize}}~q(\mathbf{i}\neq\mathbf{j})\\\nonumber
		&&\mbox{subject~to}~~q(\mathbf{i}\neq\mathbf{j})\le ~q(\mathbf{i}^\prime\neq\mathbf{j}^\prime)\\
		&&\mathbf{i}\neq\mathbf{i}^\prime~\mbox{and/or}~\mathbf{j}\neq\mathbf{j}^\prime.
	\end{eqnarray}
	Here $\mathcal{S}^{m\mapsto n}_{A/B}$ is a stochastic map mapping $m$-level probability vectors into $n$-level ones. While calculating $\mathcal{R}^{\mathfrak{C}(m)}_{\max}(n)$, for $m=2$, Lemma \ref{lemma1} allows us to restrict the optimization over the edge $\mathcal{E}_{[\alpha]}(2)$, instead of the full two-$2$-coin state space $\mathfrak{C}(2)$. In Table \ref{tab1} we list maximum payoffs for a few cases. There we also provide the optimal coin states of $\mathfrak{C}(m)$ and the applied free operations on it that maximize $\mathcal{R}^{\mathfrak{C}(m)}_{\max}(n)$. Our next result establishes quantum advantage of shared randomness generation in non-monopolize social subsidy game.
	\begin{theorem}\label{theo2}
		The optimum payoff $\mathcal{R}(n)$ for $n=3,4$ can be obtained with a coin state from $\mathfrak{Q}(2)$.
	\end{theorem}
	\begin{proof}
		Let the two-2-quoin state $\mathcal{Q}_{\mathtt{singlet}}(2):=|\psi^-\rangle_{AB}=\frac{1}{\sqrt{2}}\left(|01\rangle_{AB}-|10\rangle_{AB}\right)$ is shared between the players.	
		%Let the players share the two-2-quoin state $\mathcal{Q}_{\mathtt{singlet}}(2):=|\psi^-\rangle_{AB}=\frac{1}{\sqrt{2}}\left(|01\rangle_{AB}-|10\rangle_{AB}\right)$.
		Both of them perform the same three outcome Trine-POVM $\mathcal{M}^{\mathrm{T}}\equiv\left\lbrace\Pi_k:=\frac{2}{3}|\psi_k\rangle\langle\psi_k|\right\rbrace $, where $|\psi_k\rangle:=\cos(k-1)\theta_3|0\rangle+\sin(k-1)\theta_3|1\rangle$; $k\in\{1,2,3\},~\theta_3=2\pi/3$. This strategy leads to the coin state $\mathcal{C}^{eq}_{\ne\alpha}(3)$ yielding the optimum payoff in $\mathbb{G}(3)$. To obtain the optimum payoff in $\mathbb{G}(4)$, they consider the SIC-measurement $\mathcal{M}^{\mathrm{S}}\equiv\{\frac{1}{2}|0\rangle\langle0|, \frac{1}{2}|\psi_{k}\rangle\langle\psi_{k}|~|~ k=0,1,2\}$, where $|\psi_{k}\rangle=\sqrt{\frac{1}{3}}|0\rangle+ e^{i \frac{2k\pi}{3}}\sqrt{\frac{2}{3}}|1\rangle$. This leads to the coin state $\mathcal{C}^{eq}_{\ne\alpha}(4)$, resulting in the optimum payoff in $\mathbb{G}(4)$. This completes the proof.
	\end{proof}
	\begin{table}[t!]
		\centering
		\caption{Maximum payoff in $\mathbb{G}(n)$ given a coin state from $\mathfrak{C}(m)$. Coin states $\mathcal{C}(m)$ and the free operations $\mathcal{S}^{m\mapsto n}_A\otimes\mathcal{S}^{m\mapsto n}_B$ yielding maximum success $\mathcal{R}^{\mathfrak{C}(m)}_{\max}(n)$ are not unique in general. $\mathcal{R}^{\max}(n)$ is the maximum payoff of the game $\mathbb{G}(n)$ achievable if there is no limitation on the amount of shared randomness.}\label{tab1}
		\begin{tabular}{|c|c|c|c|c|}
			\hline 
			$~\mathcal{R}^{\mathfrak{C}(m)}_{\max}(n)~$& $~\mathcal{C}(m)~$ & $~\mathcal{S}^{m\mapsto n}_A~$ & $~\mathcal{S}^{m\mapsto n}_B~$ & $\mathcal{R}^{\max}(n)$\\ 
			\hline\hline
			& & & & \\
			$\mathcal{R}^{\mathfrak{C}(2)}_{\max}(3)=\frac{1}{8}$	& $\mathcal{C}_{1/2}(2)$ & $\begin{pmatrix}0 & \frac{1}{2}\\\frac{1}{2} & 0 \\\frac{1}{2} & \frac{1}{2}  \end{pmatrix}$ & $\begin{pmatrix}\frac{1}{2} & 0 \\0 & \frac{1}{2} \\\frac{1}{2} & \frac{1}{2} \end{pmatrix}$ & $\frac{1}{6}$ \\
			& & & & \\
			\hline
			& & & & \\ 
			$\mathcal{R}^{\mathfrak{C}(2)}_{\max}(4)=\frac{1}{15}$	& $\mathcal{C}_{1/2}(2)$ & $\begin{pmatrix}\frac{1}{5} & \frac{1}{3} \\\frac{1}{5} & \frac{1}{3} \\\frac{2}{5} & 0 \\\frac{1}{5} & \frac{1}{3}  \end{pmatrix}$ & $\begin{pmatrix}\frac{1}{3} & \frac{1}{5} \\\frac{1}{3} & \frac{1}{5} \\0 & \frac{2}{5} \\\frac{1}{3} & \frac{1}{5}  \end{pmatrix}$ & $\frac{1}{12}$\\
			& & & & \\ 
			\hline 
			& & & & \\
			$\mathcal{R}^{\mathfrak{C}(3)}_{\max}(4)=\frac{2}{27}$	& $\mathcal{C}^{eq}_{\neq\alpha}(3)$ & $\begin{pmatrix}0 & \frac{2}{3} & 0\\0 & 0 & \frac{2}{3} \\\frac{2}{3} & 0 & 0\\\frac{1}{3} & \frac{1}{3} & \frac{1}{3}  \end{pmatrix}$ & $\begin{pmatrix}0 & \frac{2}{3} & 0\\0 & 0 & \frac{2}{3} \\\frac{2}{3} & 0 & 0\\\frac{1}{3} & \frac{1}{3} & \frac{1}{3}  \end{pmatrix}$ & $\frac{1}{12}$\\ 
			& & & & \\
			\hline 
		\end{tabular}
	\end{table}
	\begin{remark}\label{remark1}
		Theorem \ref{theo1} and Theorem \ref{theo2} together provide a proof for the second part of the Proposition \ref{prop1} for $d=3,4$. According to Theorem \ref{theo1}, $\mathcal{C}^{eq}_{\ne\alpha}(d)\notin\mathfrak{S}_C(2\mapsto d)$ whenever $d>2$. In fact, from Lemma \ref{lemma2} we can say that $\mathcal{C}_{\ne\alpha}(d)\notin\mathfrak{S}_C(2\mapsto d)$. On the other hand, Theorem \ref{theo2} tells that $\mathcal{C}^{eq}_{\ne\alpha}(d)\in\mathfrak{S}_Q(2\mapsto d)$ for $d=3,4$ and hence proves the second part of the Proposition \ref{prop1}. For higher values of $d$, consider the two-2-quoin state $\mathcal{Q}_{\mathtt{singlet}}(2)$ and consider the same $d$ outcome POVM $\mathcal{M}^{(d)}\equiv\left\lbrace\Pi_k:=\frac{2}{d}|\psi_k\rangle\langle\psi_k|\right\rbrace $ for Alice and Bob, where $|\psi_k\rangle:=\cos(k-1)\theta_d|0\rangle+\sin(k-1)\theta_d|1\rangle$; $k\in\{1,\cdots,d\}$ and $\theta_d=2\pi/d$. This leads to a state $\mathcal{C}_{\ne\alpha}(d)$ and completes the proof of Proposition \ref{prop1} for arbitrary $d>2$.
	\end{remark}
	Note that for higher $d$ values the state $\mathcal{C}^{eq}_{\ne\alpha}(d)$ can not be obtained from $\mathcal{Q}_{\mathtt{singlet}}(2)$ and hence perfect payoff in $\mathbb{G}(d)$ can not be obtained even when a coin state from $\mathfrak{Q}(2)$ is given as an assistance. Optimal classical vs quantum payoff(s) for the generic case $\mathbb{G}(d)$, we leave here as an open question. 
	\subsection{Noise-robust quantum advantage} The quantum advantage established above considers a two-qubit perfect entangled state. However, entanglement is extremely fragile under noise and hence the advantage obtained with such a perfect state seems impossible to archive in a practical scenario. Thus, a more realistic question is whether the advantage manifested by quantum systems is robust to noise or not. To this aim, let us consider a noisy two-2-quoin $\mathcal{Q}_p(2):=p|\psi^-\rangle_{AB}\langle\psi^-|+(1-p)\frac{\mathbb{I}}{2}\otimes\frac{\mathbb{I}}{2}$ (i.e., a mixture of the singlet state and white noise)\footnote{Here we deal with a subclass ($p\in[0,1]$) of Werner states, for which $-\frac{1}{3}\leq p\leq1$}. Note that $\mathcal{R}^{\mathfrak{C}(2)}_{\max}(3)=1/8$ and $\mathcal{R}^{\mathfrak{C}(2)}_{\max}(4)=1/15$ (see  Table \ref{tab1}). Any quantum strategy providing a greater payoff can be considered advantageous over the classical resources. With the two-2-quoin state $\mathcal{Q}_p(2)$, if we follow the same strategy as discussed in Theorem \ref{theo2}, they can be used to demonstrate advantage over the classical strategies for $p>1/4$ and $p>1/5$ in the games $\mathbb{G}(3)$ and $\mathbb{G}(4)$. At this point, it is noteworthy that the state $\mathcal{Q}_p(2)$ is not even entangled whenever $p\le1/3$. This raises another fundamentally important question: which quantum feature does underpin the aforementioned advantage in shared randomness generation? Next, we make an attempt to provide a partial answer to this question. Recall that, a bipartite state $\rho_{AB}\in\mathcal{D}(\mathcal{H}_A\otimes\mathcal{H}_B)$ is called classically correlated (CC) if it has a diagonal representation in some orthogonal product basis, {\it i.e.} $\rho_{AB}=\sum_{a,b}p_{ab}|a\rangle_A\langle a|\otimes|b\rangle_B\langle b|$, where $\{|a\rangle_A\}$ is an orthogonal basis for $\mathcal{H}_A$ \& $\{|b\rangle_B\}$ for $\mathcal{H}_B$ and $p_{ab}\ge0,~\&~\sum_{ab}p_{ab}=1$. These states do \textit{not} possess quantum discord -- a non-classical feature present in the correlation of bipartite quantum states \cite{Ollivier2001,Henderson2001}. Besides the entangled, all the separable states, which are not CC, exhibit non-zero quantum discord. For instance, the two-2-quoin $\mathcal{Q}_p(2)$ has non-zero quantum discord for $p\in(0,1]$, whereas it is separable whenever $p\le1/3$.
	\begin{theorem}\label{theo3}
		Classically correlated bipartite quantum states will not provide any advantage in shared randomness generation.
	\end{theorem}  
	\begin{proof}
		Without loss of any generality, we can consider the computational basis and hence can represent a two-qubit CC state as $\rho_{AB}=\sum_{u,v}p_{uv}~|uv\rangle_{AB}\langle uv|;~u,v\in\{0,1\},~p_{uv}\ge0~\&~\sum_{u,v}p_{uv}=1$. To obtain a two-d-coin state $\mathcal{C}(d)$, Alice and Bob perform some $d$-outcome POVMs  $\{\mathcal{M}_{i}^{A}|\sum_{i=1}^{d}\mathcal{M}_{i}^{A}=\mathbb{I}\}$ and $\{\mathcal{M}_{i}^{B}|\sum_{i=1}^{d}\mathcal{M}_{i}^{B}=\mathbb{I}\}$ on their respective subsystems. Probability of clicking the POVM elements $\mathcal{M}_{i}^{A}\otimes\mathcal{M}_{j}^{B}$ on the state $\rho_{AB}$ is given by,
		\begin{eqnarray}
			p(ij)&=&\sum_{u,v=0}^1p_{uv}~\langle u|\mathcal{M}_{i}^{A}|u\rangle\langle v|\mathcal{M}_{j}^{B}|v\rangle.
		\end{eqnarray}
		Obviously, $\langle\psi|\mathcal{M}_{i}^X|\psi\rangle\geq0,~\&~\sum_{i=1}^{d}\langle\psi|\mathcal{M}_{i}^X|\psi\rangle=1,~\forall~|\psi\rangle;$ $X\in\{A,B\}$. This fact leads us to construct  stochastic matrices $\mathcal{S}_{X}^{2\to d}$, with the elements,
		\begin{equation}
			\left[\mathcal{S}_X^{2\to d}\right]_{kl}=\langle l|\mathcal{M}_{k}^X|l\rangle, \text{where}~ l\in\{0,1\}.
		\end{equation}
		Evidently, action of $\mathcal{S}_{A}^{2\to d}\otimes\mathcal{S}_{B}^{2\to d}$ on a classical coin $\mathcal{C}_{p}(2)\equiv p_{00}~\mathtt{hh}+p_{01}~\mathtt{ht}+p_{10}~\mathtt{th}+p_{11}~\mathtt{tt}$ will produce the same probability statistics of Eq.[3]. Therefore, any $\mathcal{C}(d)$ coin state generated from any \textit{zero-discord} two-qubit state, can be freely simulated from a properly chosen $\mathcal{C}(2)$ coin state. Hence, the quantum advantage in shared randomness generation necessarily requires the $\mathcal{Q}(2)$ states to have non-zero discord.
	\end{proof}
	
	The above theorem is quite important, as it establishes a fundamentally new application of quantum discord \cite{Ollivier2001,Henderson2001}.  Notably, several other results have been derived to establish a connection between quantum discord and entanglement transformations \cite{Cavalcanti2011, Madhok2011, Streltsov2011}, coherence resources \cite{Madhok2011a}, remote state preparations \cite{Dakic2012}, random access codes \cite{Bobby2014} etc. Our result finds a  utility of quantum discord in the generation of shared randomness.  It would be interesting to explore further quantitative connections between the measure of discord with the quantum advantage obtained in the noisy scenario. The noisy scenario becomes even more interesting if we consider distribution of sources (classical or quantum) to establish shared randomness between two parties.    
	\subsubsection{Quantum advantage in distributing SR} In the non-monopolize social subsidy game, we have considered that both the shared randomness and the strategies of the players are assisted by the referee. Let us consider now a scenario where a paired-coin (bipartite state) is prepared by one of the players who wish to distribute, through some communication channel, its one part to the other player to maximize their payoff. Distribution of the coin state $\mathcal{C}_{1/2}(2)$ in its exact form requires a perfect binary channel of capacity $1$-bit. In quantum scenario, a communication channel can be most generally described by completely positive trace preserving maps \cite{Kraus1983}. Let Alice prepare a two-$2$-quoin state $\mathcal{Q}(2)=\rho_{AB}\in\mathcal{D}(\mathbb{C}^2\otimes\mathbb{C}^2)$ in her laboratory and then she sends the $B$ part to Bob through a qubit channel $\Lambda$. They end up with a two-$2$-quoin state $\mathcal{Q}^{\prime}(2)=\mathbf{I}\otimes\Lambda\left[ \rho_{AB}\right]$, where $\mathbf{I}$ denotes the identity map ({\it i.e.} noiseless process) on the $A$ part. Then they obtain a classical shared coin from $\mathcal{Q}^{\prime}(d)$ by applying allowed free operations as suggested by the referee. In the following, we analyze two familiar noisy qubit channels for achieving better payoff in  $\mathbb{G}(n)$.
	
	{\it Qubit phase-flip channel:} Its action on an arbitrary state $\rho\in\mathcal{D}(\mathbb{C}^2)$ is given by, $\Lambda^z_p(\rho):=p~\rho+(1-p)~\sigma_z\rho\sigma_z$, where $p\in[0,1]$. If Alice sends one part of the coin state $\mathcal{Q}_{\psi^-}(2)=|\psi^-\rangle\langle\psi^-|$ to Bob through $\Lambda^z_p$ then they end up sharing the state $\mathcal{Q}^z_p(2)=\mathbf{I}_2\otimes\Lambda^z_p\left[\mathcal{Q}_{\psi^-}(2)\right]=p~\mathcal{Q}_{\psi^-}(2)+(1-p)~\mathcal{Q}_{\psi^+}(2)$. Applying the same Trine-POVM as in Theorem \ref{theo2}, the probabilities $p(ij)=\mbox{Tr}\left[\left(\Pi_i\otimes\Pi_j\right).\mathcal{Q}^z_p(2) \right]$ can be represented as a $3\times3$ matrix
	\begin{equation*}
		\mathcal{P}^z_p(3)\equiv
		\begin{pmatrix} 
			0 & \mu & \mu \\
			\mu & (1-p)~\mu & p~\mu \\
			\mu & p~\mu & (1-p)~\mu
		\end{pmatrix},
	\end{equation*}
	where $\mu=1/6$. Since the maximum payoff for $\mathbb{G}(3)$ with a perfect classical channel is $1/8$, the $\Lambda^z_p$ channel is advantageous whenever $\beta>3/4$. For the $\mathbb{G}(4)$ case, following the same SIC-POVM strategy we have, 
	\begin{equation*}
		\mathcal{P}^z_p(4)\equiv
		\begin{pmatrix} 
			0 & 3\nu &  3\nu& 3\nu \\
			3\nu & \nu^\prime &  (1+2p)\nu& (1+2p)\nu \\
			3\nu &  (1+2p)\nu & \nu^\prime&  (1+2p)\nu\\
			3\nu&  (1+2p)\nu &  (1+2p)\nu&\nu^\prime
		\end{pmatrix},
	\end{equation*}
	where $\nu=1/36$ and $\nu^\prime=(1-p)/9$. In this case maximum classical payoff is $1/15$ which means $\Lambda^z_p$ is advantageous for cases with more noise, {\it i.e.} whenever $p>7/10$. 
	
	{\it Qubit depolarizing channel:} Its action is given by, $\Lambda^D_p(\rho):=p~\rho+(1-p)~\frac{\mathbb{I}}{2}$. If Alice prepares $\mathcal{Q}_{\psi^-}(2)$, then they end up sharing the state $\mathcal{Q}^D_p(2)=p~\mathcal{Q}_{\psi^-}(2)+(1-p)~\frac{\mathbf{I}}{2}\otimes\frac{\mathbf{I}}{2}$. A straightforward calculation, in this case, yields
	\begin{equation*}
		\mathcal{P}^D_p(3)= 
		\begin{pmatrix} 
			\eta & \eta^\prime & \eta^\prime \\
			\eta^\prime & \eta & \eta^\prime \\
			\eta^\prime & \eta^\prime & \eta
		\end{pmatrix}~~\&~~\mathcal{P}^D_p(4)=
		\begin{pmatrix} 
			\delta & \delta^\prime &  \delta^\prime& \delta^\prime \\
			\delta^\prime & \delta &  \delta^\prime& \delta^\prime \\
			\delta^\prime & \delta^\prime & \delta&  \delta^\prime\\
			\delta^\prime & \delta^\prime & \delta^\prime&\delta
		\end{pmatrix},
	\end{equation*}
	where $\eta=(1-p)/9$, $\eta^\prime=(2+p)/18$, $\delta=(1-p)/16$. and $\delta^\prime=(3+p)/48$. Therefore, the channel $\Lambda^D_p$ is advantageous in $\mathbb{G}(3)$ [$\mathbb{G}(4)$] whenever $p>1/4$ [$p>1/5$]. Importantly, $\Lambda^D_p$ is an entanglement breaking channel whenever $p\le1/3$ \cite{Werner1989}. Therefore, the channel exhibits advantage in shared randomness distribution even when its quantum capacity is zero \cite{Lloyd1997,Shor2002,Devetak2005}. Also recall that classical capacity of qubit depolarizing channel is given by $\chi(\Lambda^D_p)=1-H\left(\frac{1+p}{2}\right)$, where $H(x):=-x\log_2x-(1-x)\log_2(1-x)$ is the binary entropy \cite{King2003}. Therefore, quantum advantage is tangible even when the classical capacity of the quantum channel is much less than unity.
	\section{Discussion}
	Considerable effort has been made by researchers in quantum information and foundations community to identify a list of practical instances where application of quantum rules provide advantage over the classical physics. The present work, where we establish quantum advantage in generating higher degrees of shared randomness quantified within a suitably formulated resource theoretic framework, is an addition to this list. We also show precedence of quantum channel over its classical counterpart in distributing shared randomness. Such advantage is quite relevant if we recall some of the fundamental no-go results that limit the advantage of using quantum systems in classical information processing. For instance, Holevo's theorem limits the classical capacity of a quantum channel \cite{Holevo1973} whereas the recent no-go result by Frenkel and Weiner \cite{Frenkel2015} limits classical information storage capacity in a quantum system. However, in the present work, it is established that a class of noisy qubit channels with imperfect classical capacity can surpass noiseless classical channels in distributing shared randomness. Moreover, the quantum advantage turns out to be robust to extreme noise that can erase its most prominent quantum signature, the quantum capacity. A discussion of our Proposition \ref{prop1} in connection with the seminal Bell's theorem is worth mentioning. Recall that Bell's theorem establishes a non-classical feature for quantum correlations, in the sense that some of these cannot have a classical (local realistic) description \cite{Bell66}. In the same spirit, our results also point out a non-classical feature of quantum correlations in a setting where appropriate albeit fixed measurements are performed locally on a shared bipartite state. Precisely, a two-$2$-quoin can yield correlated random variables that cannot be obtained from a two-$2$-coin. Note that, while Bell's theorem involves more than one measurement on each part of the spatially separated systems and hence requires the assumption of `measurement independence' \cite{Hall11,Barrett11}, our result only invokes a fixed measurement on each side and thus is free of this particular assumption. On the other hand, unlike Bell's theorem, the depiction of non-classical correlations in the present work requires the local dimension of the systems to be known.  
	%a restricted sense that, the obtained correlations $\mathcal{C}_{\neq\alpha}^{eq}(3)$ or, $\mathcal{C}_{\neq\alpha}^{eq}(4)$ can not be simulated from the two-level shared classical randomness, while Bell's theorem deals with subsystems of arbitrary dimension.  
	
	Our work raises a number of important questions regarding the utility of non-classical origin of randomness, which will be of interest to the broader community of researchers in quantum foundations and quantum information. First, a class of monotones, completely characterizing the (im)possibility of conversion between two shared randomness resources, is still missing. Second, the advantage of two-2-quoins and noisy qubit channels in the generation and distribution of higher level shared randomness, demonstrated in this work, necessitate further characterization of quantum resources providing such preeminence. Our work serves as a stepping stone towards unveiling the rich potentiality of accomplishing quantum advantage in shared randomness generation from higher level systems and multipartite scenarios.
	
	{\bf Acknowledgement:} We gratefully acknowledge the discussions with Guruprasad Kar. MA acknowledges support from the CSIR project 09/093(0170)/2016- EMR-I. SR acknowledges support by the Foundation for Polish Science (IRAP project, ICTQT, contract no. MAB/2018/5, co-financed by EU within Smart Growth Operational Programme). SSB is supported by the National Natural Science Foundation of China through grant 11675136, the Foundational Questions Institute through grant FQXiRFP3-1325, the Hong Kong Research Grant Council through grant 17300918, and the John Templeton Foundation through grant 60609, Quantum Causal Structures. The opinions expressed in this publication are those of the authors and do not necessarily reflect the views of the John Templeton Foundation. MA  and
	MB acknowledge support through the research grant of
	INSPIRE Faculty fellowship of MB from the Department
	of Science and Technology, Government of India.
	
	%\showacknow{} % Display the acknowledgments section
	

\end{document}